\documentclass[prd,preprintnumbers,nofootinbib,twocolumn,superscriptaddress]{revtex4} % Physical Review D
\usepackage{graphics,graphicx}
\usepackage{amsfonts,amsmath,amsthm,amssymb}

\begin{document}

\preprint{IPMU11-0155}

\title{No Classicalization Beyond Spherical Symmetry}

\author{Ratindranath Akhoury}
\email{akhoury@umich.edu}
\affiliation{Michigan Center for Theoretical Physics, Randall Laboratory of Physics, University of Michigan, Ann Arbor, MI 48109-1120, USA}

\author{Shinji Mukohyama}
\email{shinji.mukohyama@ipmu.jp}
\affiliation{IPMU, The University of Tokyo, Kashiwa, Chiba 277-8582, Japan}

\author{Ryo Saotome}
\email{rsaotome@umich.edu}
\affiliation{Michigan Center for Theoretical Physics, Randall Laboratory of Physics, University of Michigan, Ann Arbor, MI 48109-1120, USA}

\date{\today}

\begin{abstract}
 We point out that a field theory that exhibits the classicalization
 phenomenon for perfect spherical symmetry ceases to do so when the
 spherical symmetry is significantly relaxed. We first investigate a small
 non-spherical deformation and show that the classicalization radius
 tends to decrease in a region where a shell made of the field is
 slightly flattened. Next, in order to describe a sufficiently large
 flattened region, we consider a high-energy collision of planar shells
 and show that the system never classicalizes before reaching sub-cutoff
 lengths. This no-go result is further strengthened by an analysis of a small
 non-planar deformation. Finally, we show that the shape of a scattered
 planar wave is UV sensitive. 
\end{abstract}

\maketitle

\section{Introduction}

Recently a novel approach to the UV-completion of a class of
non-renormalizable theories was proposed~\cite{Dvali:2010jz}. In this
approach, called {\it classicalization}, high-energy scattering
amplitudes are conjectured to be unitarized by the production of extended
classical objects, dubbed {\it classicalons}. Although the existence and
properties of classicalons have not been studied at all, it has been
somehow suspected that they might form below the length scale called
the {\it classicalization radius}~\cite{Dvali:2010ns,Dvali:2011th}. The
classicalization radius is defined as the classical radius down to which
high-energy shells of fields propagate essentially freely, without
experiencing a significant correction from interaction. We can at least
consider the condition 
\begin{equation}
 (\mbox{classicalization radius}) \gg 
  (\mbox{cutoff length}) \label{eqn:classicalization-condition}
\end{equation}
as a necessary condition for the formation of classicalons.

If classicalization really works then we would have to change our
view of non-renormalizable field theories and Wilsonian UV completion,
in which a weakly coupled quantum field theory above the cutoff scale of
an effective field theory is reconstructed by integrating-in some new
degrees of freedom. In the classicalization approach, a
non-renormalizable field theory would unitarize itself by its own
resources, above the scale that would otherwise be the UV cutoff of the
theory.

Hence it is important to investigate the validity of the classicalization
proposal. However, so far, the studies of the classicalization approach have been
rather limited. One of the limitations is that all considerations in the 
literature assume perfect spherical symmetry. The purpose of this paper
is to study the classicalization proposal beyond spherical
symmetry. Unfortunately, we shall see that the above mentioned necessary 
condition (\ref{eqn:classicalization-condition}) does not hold if
spherical symmetry is significantly relaxed. Thus, considerations in the
present paper point towards the conclusion that classicalization does
not serve as UV-completion for the class of non-renormalizable
theories.

The rest of the paper is organized as follows. In Sec.~\ref{sec:review}
we shall briefly review the classicalization proposal. We shall also
argue that classicalization does not serve as a way to UV complete
the other classes of the field theories considered in the literature, except for the   
Goldstone-type field theory. For this reason, we shall investigate the 
Goldstone-type field theory in starting with Sec.~\ref{sec:non-spherical}. 
In Sec.~\ref{sec:non-spherical} we consider a small non-spherical
deformation and show that the classicalization radius tends to decrease 
in a region where a shell made of the field is slightly flattened. In
Sec.~\ref{sec:planar-case}, in order to describe a sufficiently large
flattened region, we consider the high-energy collision of planar shells and
show that the system never classicalizes before reaching sub-cutoff
lengths. This no-go result is further strengthened by the analysis of a small
non-planar deformation in Sec.~\ref{sec:non-planar-deviation}. In
Sec.~\ref{sec:UVsensitivity} we show that the shape of scattered planar
wave is UV sensitive. Sec.~\ref{sec:conclusion} is devoted to a summary
of the main results of this paper.

Throughout this paper we shall adopt the mostly plus sign ($-+++$) for
the metric.

\section{A Brief review of the classicalization proposal}
\label{sec:review}

In the literature, the classicalization proposal has been studied for a
scalar field without shift symmetry, a scalar field with shift symmetry
(Goldstone-type field), and the graviton in general relativity. In this
paper we shall investigate the robustness of the classicalization
proposal for the theory of Goldstone bosons. Before going into further 
detail, let us explain why we will not consider the other two cases.

The example of the scalar field without shift symmetry has a Lagrangian
of the form 
\begin{equation}
 {\cal L} = -\frac{1}{2}g^{\mu\nu}
  \partial_{\mu}\phi\partial_{\nu}\phi
-\frac{\phi}{2M_*}g^{\mu\nu}
  \partial_{\mu}\phi\partial_{\nu}\phi
  + \cdots.  
\end{equation}
In ref.~\cite{Dvali:2010jz}, the second term on the right hand side was
considered as a possible source of classicalization. However, one can
redefine the field $\phi$ so that the sum of the first and the second
terms becomes a canonical kinetic term of the new field
$\phi_c=(2/3)(1+\phi/M_*)^{3/2}$. Thus, the system is equivalent to a
noninteracting canonical scalar field. The field redefinition does not
work for $\phi<-M_*$, but in this regime $\bar{\phi}_c=|\phi_c|$ has a
wrong sign kinetic term and unitarity is violated. Therefore, the first
two terms in this Lagrangian do not serve as a good example of
classicalization.

The graviton in Einstein gravity has been considered as a possible candidate
that exhibits the classicalization
phenomenon~\cite{Dvali:2011th,Dvali:2010bf}. However, it is well known that
in general relativity, one can form a black hole with arbitrarily
small mass~\cite{Choptuik:1992jv}. Therefore, in general relativity one
can in principle probe arbitrarily short distances unless new
physics above the cutoff scale somehow prevents one from doing
so. Therefore, classicalization is unlikely to work in general setups in
general relativity.

In this paper we shall consider the Goldstone-type field, i.e. a scalar
field with shift symmetry, since this is the only remaining situation that has
been studied in the literature so far.

Classicalization of the Goldstone-type field was investigated in 
the Euclidean path integral formulation in \cite{Bajc:2011ey}. However, 
the background solution is singular and the corresponding Euclidean
action is infinite. Therefore, it seems rather difficult to extract any
physically meaningful insights from the result of
\cite{Bajc:2011ey}. For this reason, in the present paper we will not
adopt the path integral formulation.

The Goldstone-type field that has been studied in the context of
classicalization has a Lagrangian of the form
\begin{equation}
{\cal L}  = -\frac{1}{2}g^{\mu\nu}
 \partial_{\mu}\phi\partial_{\nu}\phi
 + \frac{L_*^4}{4}
 \left[g^{\mu\nu}
  \partial_{\mu}\phi\partial_{\nu}\phi\right]^2.
 \label{eqn:lagrangian}
\end{equation}
Following refs.~\cite{Dvali:2010ns,Dvali:2011th}, we investigate
the classical dynamics of a spherical shell made of this scalar field in
a Minkowski spacetime. Namely, we assume that the backreaction of the
scalar field on the background geometry is negligible and we seek a
solution $\phi(t,r)$ to the equation of motion
\begin{equation}
 \Box_4 \phi =  L_*^4\partial^{\mu}
  \left[(\partial\phi)^2\partial_{\mu}\phi\right]
\end{equation}
in a Minkowski spacetime
\begin{equation}
 ds^2 = -dt^2 + dr^2 +
  r^2(d\theta^2+\sin^2(\theta)d\bar{\theta}^2). 
  \label{eqn:flat4dmetric}
\end{equation}
Our neglect of the backreaction effects  is justified in the decoupling limit
$M_{pl}L_* \gg 1$.

What we are interested in is the classical radius $r_*$ down to which
high-energy spherical shells propagate essentially freely, without
experiencing a significant correction from the interaction term. This
radius is dubbed the {\it classicalization radius}. In order to determine
$r_*$ for this system, refs.~\cite{Dvali:2010ns,Dvali:2011th} solved the
equation of motion iteratively by expanding $\phi$ as  
\begin{equation}
 \phi = \phi_0+\phi_1+\cdots,
\end{equation}
and assuming that $\phi_0$ satisfies the equation of motion with
$L_*=0$. We shall soon see what the expansion parameter is.

Under the assumed spherical symmetry, a general solution to the zeroth
order equation $\Box_4\phi_0=0$ is 
\begin{equation}
 \phi_0 = \frac{1}{r}\left[F_0(t+r)-F_0(t-r)\right],
\end{equation}
where we have imposed the regularity of $\phi_0$ at $r=0$. In order for
this solution to represent a shell with thickness $a$, the form of the
function $F_0(t)$ is set to 
\begin{equation}
 F_0(t) = A f(t/a),
\end{equation}
where $f(\tau)$ is a function whose amplitude and derivatives are of
$O(1)$ in the vicinity of the shell configuration, so that the total
energy and the occupation number of the configuration are $\sim A^2/a$
and $\sim A^2$, respectively. According to
refs.~\cite{Dvali:2010ns,Dvali:2011th}, we are interested in
configurations with a small occupation number. Thus we suppose that 
\begin{equation}
 A = O(1).
\end{equation}

We are interested in the behavior of the system before the thin shell
peaked at $t+r\sim 0$ reaches the center. Hence we can safely drop
$-F_0(t-r)$. We thus have
\begin{equation}
 \phi_0 = \frac{A}{r}f(\tau_+), \quad \tau_+ = \frac{t+r}{a}.
  \label{eqn:phi0-spherical}
\end{equation}
Since $\phi_0$ is small at large $r$ but becomes large at small $r$, the 
nonlinear interaction is negligible at large $r$ but becomes significant 
at small $r$. The classicalization radius $r_*$ for this system is thus
defined as a radius at which the amplitude of $\phi_1$ catches up with
that of $\phi_0$.

One can easily guess $r_*$ from (\ref{eqn:phi0-spherical}). Since
\begin{equation}
 g^{\mu\nu}\partial_{\mu}\phi\partial_{\nu}\phi
  = -\frac{A^2}{ar^3}\left[f'(\tau_+)f(\tau_+)+O(a/r)\right],
\end{equation}
the second term in the Lagrangian (\ref{eqn:lagrangian}) becomes as
important as the first term when 
\begin{equation}
 \frac{L_*^4}{ar^3} \sim 1. 
\end{equation}
Thus, it is expected that the classicalization radius $r_*$ is
\begin{equation}
 r_* \sim (L_*^4/a)^{1/3}.
\label{eqn:r*-spherical}
\end{equation}

To confirm this expectation, let us seek the first order solution
$\phi_1$, which is the retarded solution to 
\begin{equation}
 \Box_4 \phi_1 =  L_*^4\partial^{\mu}
  \left[(\partial\phi_0)^2\partial_{\mu}\phi_0\right].
\end{equation}
Since the right hand side is calculated as 
\begin{eqnarray}
 \bar{s}_0(\tau_+,r) & = & \frac{A^3L_*^4}{2a^2r^5}
  \left\{
   \left[f''(\tau_+)(f(\tau_+))^2
   \right.\right.\nonumber\\
 & & \left.\left.
	    +4(f'(\tau_+))^2f(\tau_+)\right]
  +O(a/r)\right\},
\end{eqnarray}
we obtain
\begin{eqnarray}
 \phi_1 & = & \frac{a}{2r}\int_r^{\infty}dr'r'
  \int_{-\infty}^{\tau_+}d\tau_1\bar{s}_0(\tau_1,r')
  \nonumber\\
 & = & 
  -\frac{A^3L_*^4}{12ar^4}\int_{-\infty}^{\tau_+}d\tau_1
  \left\{
   \left[f''(\tau_1)(f(\tau_1))^2
   \right.\right.\nonumber\\
 & & \left.\left.
	    +4(f'(\tau_1))^2f(\tau_1)\right]
 + O(a/r)\right\}.
\end{eqnarray}
Since the integral is of $O(1)$ in the vicinity of the shell, the
amplitude of $\phi_1$ catches up with that of $\phi_0$ at $r\sim r_*$,
where $r_*$ is given by (\ref{eqn:r*-spherical}).

\section{Non-spherical deformation}
\label{sec:non-spherical}

In this section we consider a non-spherical deformation of the shell and
see that if a part of the shell is flatter (or more curved), $r_*$ is 
slightly reduced (or increased) in that part.

\subsection{zeroth order solution} 

For the zeroth order part, we adopt the ansatz
\begin{equation}
 \phi_0 = \phi_{00}(t,r) + \phi_{02}(t,r)P_2(\cos\theta). 
\end{equation}
Then the zeroth order equation leads to the solution
\begin{eqnarray}
 \phi_{00} & = & \frac{1}{r}\left[F_0(t+r)+G_0(t-r)\right], \nonumber\\
 \phi_{02} & = & r\left(\partial_r \frac{1}{r}\right)^2
  \left[F_2(t+r)+G_2(t-r)\right]. 
\end{eqnarray}
Imposing regularity at $r=0$ leads to the condition
$F_0(t)+G_0(t)=F_2(t)+G_2(t)=0$ for ${}^{\forall}t$. Thus, the zeroth
order solution is 
\begin{eqnarray}
 \phi_0 & = & \frac{1}{r}\left[F_0(t+r)-F_0(t-r)\right]
  \nonumber\\
 & & + r\left(\partial_r\frac{1}{r}\right)^2
   \left[F_2(t+r)-F_2(t-r)\right] P_2(\cos\theta). 
   \label{eqn:zeroth-sol}
\end{eqnarray}

Let us suppose that 
\begin{equation}
 F_0(t) = Af(t/a), \quad
 F_2(t) = 
  Aar_2 \int_{-\infty}^{t/a}d\tau_1f(\tau_1), 
\end{equation}
where $r_2$ is a constant. For $r\gg a$, this solution is reduced 
to 
\begin{eqnarray}
 \phi_0 & \simeq &
  \frac{A}{r}
  \left\{
   \left[f(\tau_+)-f(\tau_-)\right]
   + \left[f'(\tau_+)-f'(\tau_-)\right]\frac{r_2}{a}P_2(\cos\theta)
  \right\} \nonumber\\
 & = & 
  \frac{A}{r}
  \left[f(\tilde{\tau}_+)-f(\tilde{\tau}_-)\right]
  + O(r_2^2/a^2),
  \label{eqn:phi0r>>l}
\end{eqnarray}
where
\begin{equation}
 \tau_{\pm}\equiv \frac{t\pm r}{a}, \quad
 \tilde{\tau}_{\pm} = \tau_{\pm} + \frac{r_2}{a}P_2(\cos\theta).  
  \label{eqn:def-tildetau}
\end{equation}
If $f(\tau)$ represents a wave packet peaked at $\tau=0$ then
$f(\tilde{\tau}_{\pm})$ is peaked at 
$r\simeq \mp [t+r_2P(\cos\theta)]$.  

The non-spherical deformation $\phi_{02}$ includes $O(a/r)$ and
$O(a^2/r^2)$ corrections. The $O(a/r)$ correction to
(\ref{eqn:phi0r>>l}) slightly changes the height of the peak. Indeed, by
including the $O(a/r)$ correction, (\ref{eqn:phi0r>>l}) is modified as 
\begin{eqnarray}
 \phi_0 
  & = & 
  \frac{A}{r}\left\{
 f(\tilde{\tau}_+)
  \left[1-\frac{3r_2}{r}P_2(\cos\theta)\right]
  \right.\nonumber\\
 & & 
  - f(\tilde{\tau}_+)
  \left[1+\frac{3r_2}{r}P_2(\cos\theta)\right]
  \nonumber\\
 & & \left.
  + O(a^2/r^2) + O(r_2^2/a^2)\right\},
  \label{eqn:phi0O(l/r)}
\end{eqnarray}
Although we shall not write it explicitly here, the $O(a^2/r^2)$
correction represents the tail induced by the motion of the deformed
shell.

\subsection{Retarded solution to inhomogeneous equation}

It is easy to show that 
\begin{equation}
 \phi_1(t_+,t_-) = \frac{\psi_{10}}{r}
  + r\left(\partial_r\frac{1}{r}\right)^2\psi_{12}
  P_2(\cos\theta)
  \label{eqn:retardedsol}
\end{equation}
with 
\begin{equation}
 \psi_{1n} = -\frac{1}{4}\int_{-\infty}^{t_+}dt'_+
  \int_{-\infty}^{t_-}dt'_- f_n(t'_+,t'_-) \quad (n=0,2)
  \label{eqn:psi1n}
\end{equation}
satisfies the inhomogeneous equation
\begin{equation}
 \Box_4 \phi_1 = 
  \frac{f_0(t_+,t_-)}{r}
  + r\left(\partial_r\frac{1}{r}\right)^2f_2(t_+,t_-)
  P_2(\cos\theta) 
\end{equation}
and the retarded boundary condition. Note $t_{\pm}\equiv t\pm r$.

There is freedom to add $B_2(t)r^3+C_2(t)r$ to $f_2(t_+,t_-)$ but
$B_2(t)$ and $C_2(t)$ can be fixed by demanding that $f_2(t_+,t_-)$
remains finite at $r\to\infty$ for ${}^{\forall}t$. Thus, when the
inhomogeneous equation is of the form 
\begin{equation}
 \Box_4 \phi_1 = 
  s_0(t,r) + s_2(t,r) P_2(\cos\theta),
  \label{eqn:boxphi1}
\end{equation}
we have
\begin{eqnarray}
 f_0(t_+,t_-) & = & r s_0(t,r), \nonumber\\
 f_2(t_+,t_-) & = & r\int_{r}^{\infty}dr_1 
  r_1\int_{r_1}^{\infty}dr_2\frac{s_2(t,r_2)}{r_2}, 
\end{eqnarray}
where $t$ and $r$ on the right hand side should be understood as
$(t_++t_-)/2$ and $(t_+-t_-)/2$ respectively.

\subsection{First order solution up to $O(r_2/a)$ with $O(a/r)$
  correction}

We are interested in the behavior of the system before the thin shell
peaked at $t+r\sim 0$ reaches the center. Hence, we can safely drop
$-F_0(t-r)$ and $-F_2(t-r)$ in (\ref{eqn:zeroth-sol}). We 
thus have 
\begin{eqnarray}
 \phi_0 & = &
  \frac{A}{r}
  \left\{ f(\tau_+)
   \left[1-\frac{3r_2}{r}P_2(\cos\theta)\right]
   \right.\nonumber\\
 & & \left.
   +f'(\tau_+)\frac{r_2}{a}P_2(\cos\theta)
   + O(a^2/r^2)
  \right\} \nonumber\\
 & = & \frac{A}{r}
  \left\{ f(\tilde{\tau}_+)
   \left[1-\frac{3r_2}{r}P_2(\cos\theta)\right]
   \right. \nonumber\\
 & & \left.
   + O(a^2/r^2) + O(r_2^2/a^2)\right\},
\end{eqnarray}
where $\tau_+$ and $\tilde{\tau}_+$ are defined in
(\ref{eqn:def-tildetau}). Consequently, up to $O(r_2/a)$ the equation
for $\phi_1$ is of the form (\ref{eqn:boxphi1}) with 
\begin{eqnarray}
 \bar{s}_0 & = & 
  \frac{A^3L_*^4}{2a^2r^5}
  \left\{\left[f''(\tau_+)(f(\tau_+))^2
	  +4(f'(\tau_+))^2f(\tau_+)\right]
  \right. \nonumber\\
 & & \left.
   - \frac{3l}{r}f'(\tau_+)(f(\tau_+))^2 
   + O(a^2/r^2) \right\}, \nonumber\\
 \bar{s}_2 & =& 
 \frac{r_2}{a}\left\{
  \frac{\partial \bar{s}_0}{\partial \tau_+}
  \right.\nonumber\\
 & & 
  -\frac{15A^3L_*^4}{2ar^6}
  \left[f''(\tau_+)(f(\tau_+))^2+4(f'(\tau_+))^2f(\tau_+)\right]
  \nonumber\\
 & & 
 \left. +  \frac{A^3L_*^4}{a^2r^5}\times O(a^2/r^2)
\right\}. \label{eqn:s0ands2}
\end{eqnarray}
where $\bar{s}_n\equiv s_n(t=a\tau_+-r,r)$ ($n=0,2$) is $s_n$ 
written as a function of ($\tau_+$, $r$). Note that in the coordinates
($\tau_+$, $r$, $\theta$, $\bar{\theta}$), the Minkowski metric
(\ref{eqn:flat4dmetric}) is written as
\begin{equation}
 ds^2 = -a^2d\tau_+^2 + 2ad\tau_+dr 
  + r^2(d\theta^2+\sin^2\theta d\bar{\theta}^2),
\end{equation}
and
\begin{equation}
 g^{\mu\nu}\partial_{\mu}\partial_{\nu}
  = \frac{1}{a}\partial_+\partial_r + \frac{1}{4}\partial_r^2
  + \frac{1}{r^2}\partial_{\theta}^2 
  + \frac{1}{r^2\sin^2\theta}\partial_{\bar{\theta}}^2. 
\end{equation}

From $\bar{s}_2$, $f_2$ is obtained by solving 
\begin{eqnarray}
 \bar{s}_2 & =&
  r\left[
    \left(\frac{\partial}{\partial r}
     +\frac{1}{a}\frac{\partial}{\partial\tau_+}\right)
    \frac{1}{r}\right]^2\bar{f}_2 \nonumber\\
 & = &\frac{1}{a^2r}
  \left[\frac{\partial^2}{\partial\tau_+^2}
	  - \frac{3a}{r}\frac{\partial}{\partial\tau_+}
	  + O(a^2/r^2)
		\right]\bar{f}_2.
\end{eqnarray}
Here and in the following, 
$\bar{f}_n\equiv f_n(t_+=a\tau_+,t_-=a\tau_+-2r)$
($n=0,2$) is $f_n$ written as a function of ($\tau_+$, $r$). Thus, we
obtain 
\begin{eqnarray}
 \bar{f}_2(\tau_+,r) 
  & = & a^2 r
  \left[
   \int_{-\infty}^{\tau_+}d\tau_1
   \int_{-\infty}^{\tau_1}d\tau_2
   \bar{s}_2(\tau_2,r)
   \right. \nonumber\\
 & & 
   + \frac{3a}{r}
   \int_{-\infty}^{\tau_+}d\tau_1
   \int_{-\infty}^{\tau_1}d\tau_2
   \int_{-\infty}^{\tau_2}d\tau_3
   \bar{s}_2(\tau_3,r)
   \nonumber\\
 & & 
  \left.
   + O(a^2/r^2)
  \right]. 
\end{eqnarray}
We of course have
\begin{equation}
 \bar{f}_0(\tau_+,r) = r \bar{s}_0(r,\tau_+). 
\end{equation}

Now by changing the integration variables from ($t_+'$, $t_-'$) to
($\tau_+'=t_+'/a$,$r'=(t'_+-t'_-)/2$), the formula (\ref{eqn:psi1n}) is 
written as
\begin{equation}
 \bar{\psi}_{1n}(\tau_+,r) = -\frac{a}{2}
  \int_{r}^{\infty}dr' 
  \int_{-\infty}^{\tau_+}d\tau'_+
  \bar{f}_n(\tau'_+,r') \quad (n=0,2),
\end{equation}
where 
$\bar{\psi}_{1n}\equiv \psi_{1n}(t_+=a\tau_+,t_-=a\tau_+-2r)$ 
is $\psi_{1n}$ written as a function of ($\tau_+$, $r$). Concretely, 
\begin{eqnarray}
 \bar{\psi}_{10}(\tau_+,r) & = & -\frac{a}{2}
  \int_{r}^{\infty}dr' r' \int_{-\infty}^{\tau_+}d\tau_1
  \bar{s}_0(\tau_1,r'), \nonumber\\
 \bar{\psi}_{12}(\tau_+,r) & = & -\frac{a^3}{2}
  \int_{r}^{\infty}dr' r' \nonumber\\
 & & 
  \times\left[
   \int_{-\infty}^{\tau_+}d\tau_1
   \int_{-\infty}^{\tau_1}d\tau_2 
   \int_{-\infty}^{\tau_2}d\tau_3 \bar{s}_2(\tau_3,r')
   \right.\nonumber\\
 & & 
      + \frac{3a}{r'}
   \int_{-\infty}^{\tau_+}d\tau_1
   \int_{-\infty}^{\tau_1}d\tau_2 
   \int_{-\infty}^{\tau_2}d\tau_3
   \nonumber\\
 & & \left. 
      \times\int_{-\infty}^{\tau_3}d\tau_4
   \bar{s}_2(\tau_4,r') 
   + O(a^2/{r'}^2) \right].
\end{eqnarray}

Therefore, up to $O(r_2/a)$, the formula (\ref{eqn:retardedsol}) gives 
\begin{equation}
 \phi_1 \simeq \bar{\phi}_{10}(\tau_+,r)
  + \bar{\phi}_{12}(\tau_+,r)P_2(\cos\theta), 
\end{equation}
where
\begin{eqnarray}
 \bar{\phi}_{10} & = & 
  -\frac{a}{2r}
  \int_{r}^{\infty}dr' r' \int_{-\infty}^{\tau_+}d\tau_1
  \bar{s}_0(\tau_1,r'), \nonumber\\
 \bar{\phi}_{12} & = & 
  -\frac{a}{2r}
  \left[
  \int_{r}^{\infty}dr' r' \int_{-\infty}^{\tau_+}d\tau_1
  \bar{s}_2(\tau_1,r') \right. \nonumber\\
 & & 
      - \frac{3a}{r}
  \int_{r}^{\infty}dr'  \int_{r'}^{\infty}dr''
  \int_{-\infty}^{\tau_+}d\tau_1
  \int_{-\infty}^{\tau_1}d\tau_2
  \bar{s}_2(\tau_2,r') \nonumber\\
& & \left. + O(a^2/r^2)  \right]. 
\end{eqnarray}
Here, we have integrated by parts in order to obtain the expression for 
$\bar{\phi}_{12}$. Finally, by using the explicit expressions
(\ref{eqn:s0ands2}), it is shown that 
\begin{eqnarray}
 \bar{\phi}_{10} & = & 
  -\frac{A^3L_*^4}{12ar^4}
  \left\{ g(\tau_+)
   - \frac{3a}{4r}(f(\tau_+))^3 + O(a^2/r^2)\right\},
  \nonumber\\
 \bar{\phi}_{12} & = & 
  -\frac{A^3L_*^4}{12ar^4}\frac{r_2}{a}
  \left\{ g'(\tau_+)
   - \frac{9a}{4r}f'(\tau_+)(f(\tau_+))^2
   \right. \nonumber\\
 & & 
  \left.
   - \frac{12a}{r} g(\tau_+)
	 \right\},
\end{eqnarray}
where
\begin{equation}
 g(\tau_+) \equiv
   \int_{-\infty}^{\tau_+}d\tau_1
   \left[f''(\tau_1)(f(\tau_1))^2+4(f'(\tau_1))^2f(\tau_1)\right]. 
\end{equation}

In summary, 
\begin{eqnarray}
 \phi_1 & =& 
  -\frac{A^3L_*^4}{12ar^4}
  \left\{g(\tilde{\tau}_+)
   \left[1-\frac{12r_2}{r}P_2(\cos\theta)\right]
- \frac{3a}{4r}(f(\tilde{\tau}_+))^3 \right.\nonumber\\
 & & \left. + O(a^2/r^2) + O(r_2^2/a^2)\right\}.
\end{eqnarray}
Hence, we obtain
\begin{eqnarray}
\frac{\phi_1}{\phi_0} & = & 
 -\frac{A^2L_*^4}{12ar^3}
 \left\{   \frac{g(\tilde{\tau}_+)}{f(\tilde{\tau}_+)}
  \left[1-\frac{9r_2}{r}P_2(\cos\theta)\right]
  - \frac{3a}{4r}(f(\tilde{\tau}_+))^2 \right.\nonumber\\
 & & \left. + O(a^2/r^2) + O(r_2^2/a^2)\right\}. 
\end{eqnarray}
As noted earlier, the classicalization radius $r_*$ is the value of $r$ at which the
amplitude of $\phi_1$ catches up with that of $\phi_0$. Thus, this
result clearly shows that $r_*$ is slightly reduced (or increased) if 
$r_2P_2(\cos\theta)$ is positive (or negative), i.e. if the shell is
flatter (or more curved) in some region. This is exactly what we have
expected.

In the next section, in order to describe a sufficiently large flattened
region, we consider a planarly symmetric case.

\section{Planarly symmetric case}
\label{sec:planar-case}

In $4$-dimensional flat spacetime
\begin{equation}
 ds_4^2 = -dt^2+dx^2+dy^2+dz^2,
\end{equation}
we consider a planarly symmetric ansatz $\phi=\phi(t,x)$. By integrating
out the two spatial dimensions parallel to the plane of symmetry, $y$
and $z$, we obtain a scalar field $\varphi=\sqrt{\mathcal{A}_2}\phi$ in 
$2$-dimension, where $\mathcal{A}_2$ is the area of the plane.

In a two dimensional flat spacetime
\begin{equation}
 ds_2^2 = -dt^2+dx^2 = -dt_+dt_-, \quad t_{\pm} \equiv t\pm x,
\end{equation}
the retarded solution to 
\begin{equation}
\Box_2 \varphi_1 = f(t_+,t_-)
\end{equation}
is given by
\begin{equation}
 \varphi_1 = -\frac{1}{4}\int_{-\infty}^{t_+}dt'_+
  \int_{-\infty}^{t_-}dt'_- f(t'_+,t'_-). 
\end{equation}

The equation of motion 
\begin{equation}
 \Box_4 \phi =  L_*^4\partial^{\mu}
  \left[(\partial\phi)^2\partial_{\mu}\phi\right]
\end{equation}
in $4$-dimensions is reduced to 
\begin{equation}
 \Box_2 \varphi =  \frac{L_*^4}{\mathcal{A}_2}\partial^{\mu}
  \left[(\partial\varphi)^2\partial_{\mu}\varphi\right]
\end{equation}
in $2$-dimensions. Let us solve this equation iteratively by expanding
$\varphi$ as 
\begin{equation}
 \varphi = \varphi_0+\varphi_1+\cdots,
\end{equation}
and assuming that $\varphi_0$ satisfies the equation of motion with
$L_*=0$, i.e., $\Box_2 \varphi = 0$. Thus, 
\begin{equation}
 \varphi_0 = \varphi_+(t_+) + \varphi_-(t_-). 
\end{equation}
We shall soon see what the expansion parameter is.

The first order equation is
\begin{equation}
 \Box_2\varphi_1 = \frac{8L_*^4}{\mathcal{A}_2}
  \left[ \varphi_+''(t_+)(\varphi_-'(t_-))^2
   +\varphi_-(t_-)''(\varphi_+'(t_+))^2\right]. 
\end{equation}
Thus the retarded solution is
\begin{eqnarray}
 \varphi_1 & = & -\frac{2L_*^4}{\mathcal{A}_2}\int_{-\infty}^{t_+}dt'_+
  \int_{-\infty}^{t_-}dt'_- 
  \left[ \varphi_+''(t_+')(\varphi_-'(t_-'))^2 \right.\nonumber\\
 & & \left.
   +\varphi_-(t_-')''(\varphi_+'(t_+'))^2\right]. 
\end{eqnarray}

Let us suppose that
\begin{equation}
 \varphi_+(t)=\varphi_-(t) = A\ f(t/a). 
\end{equation}
Roughly speaking, the total energy and the occupation number of the 
configuration are $\sim A^2/a$ and $\sim A^2$. Then we obtain 
\begin{eqnarray}
 \frac{\varphi_1}{A} & = & -\frac{2L_*^4A^2}{\mathcal{A}_2\, a^2}
  \int_{-\infty}^{t_+/a}d\tau_1
  \int_{-\infty}^{t_-/a}d\tau_2 
  \left[ f''(\tau_1)(f'(\tau_2))^2 \right.\nonumber\\
 & & \left. +f''(\tau_2)(f'(\tau_1))^2\right]. 
\end{eqnarray}
We now see that the expansion parameter is
\begin{equation}
 \epsilon \equiv \frac{2L_*^4A^2}{\mathcal{A}_2\, a^2}.
\end{equation}
The iterative solution is a good approximation if $\epsilon\ll 1$.

\begin{figure}[htb]
\begin{center}
\includegraphics[width=8.0cm]{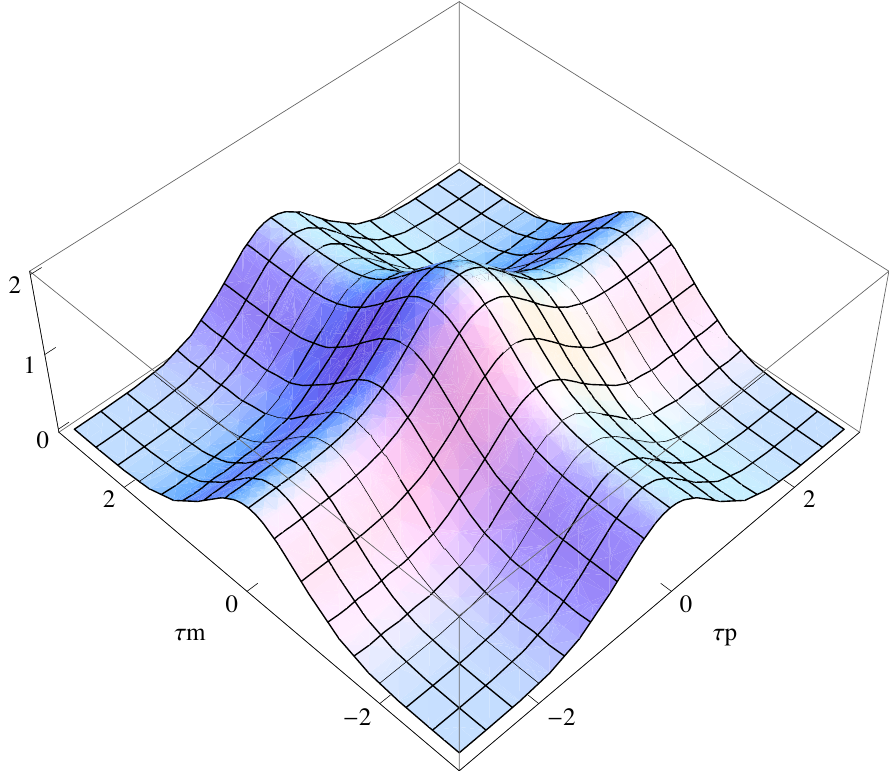}
\caption{The plot of $\varphi_0/A$ as a function of 
$\tau_{\pm}\equiv t_{\pm}/a$ for $f(\tau)=e^{-\tau^2}$.
} 
\label{fig:phi0}
\end{center}
\end{figure}

\begin{figure}[htb]
\begin{center}
\includegraphics[width=8.0cm]{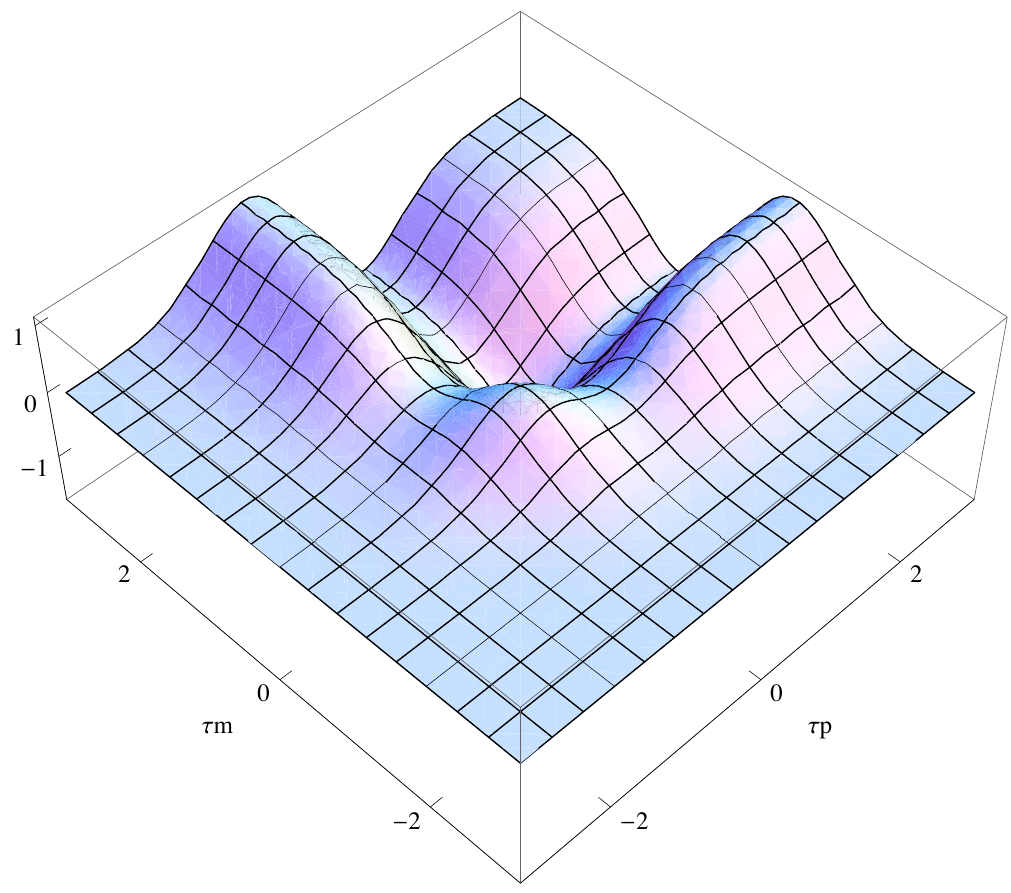}
\caption{The plot of $\varphi_1/(-A\epsilon)$ as a function of 
$\tau_{\pm}\equiv t_{\pm}/a$ for $f(\tau)=e^{-\tau^2}$.
} 
\label{fig:phi1}
\end{center}
\end{figure}

As an example, let us consider a Gaussian wave packet
\begin{equation}
 f(\tau) = e^{-\tau^2}. 
\end{equation}
Figs~\ref{fig:phi0} and \ref{fig:phi1} show $\varphi_0/A$ and
$\varphi_1/(-A\epsilon)$, respectively, as functions of 
$\tau_{\pm}\equiv t_{\pm}/a$. This clearly shows that $\varphi_1$ 
stays small compared to $\varphi_0$ as far as $\epsilon\ll 1$ and that
the interaction length is $\sim a$. In this case the system never
classicalizes. For $\epsilon\sim 1$, the amplitude of $\varphi_1$
becomes as large as that of $\varphi_0$ but the interaction length is
still $\sim a$. Thus, for high-energy scattering with $a\ll L_*$, there
is no sign of classicalization before reaching the sub-cutoff length.

\section{Deviation from perfect planar symmetry}
\label{sec:non-planar-deviation}

In this section we will consider deviations from the perfect planar
symmetry discussed above. Our purpose in doing so is to see if the
fluctuations will introduce a classicalization radius. Since
classicalization was shown for spherical symmetry, our expectation is to
observe a tendency away from the results of the strictly planar
symmetric case. 

We want to probe in a 4-dimensional spacetime,  the behavior of the field:
\begin{align}
\phi(t,x,y,z)=\phi_{0}(t,x,y,z)+\phi_{1}(t,x,y,z)+... \, \, ,
\end{align}
where $\phi_{0}$ satisfies $\Box_{4}\phi_{0}=0$. In terms of the rescaled fields $\varphi$ of the previous section, let us make the ansatz:
\begin{align}
\phi_{0}(t,x,y,z)&=\frac{1}{\sqrt {\mathcal {A}}_2}\varphi_{00}(t,x)
\nonumber \\
&+\frac{1}{\sqrt {\mathcal {A}}_2}\varphi_{01}(t,x) Re(e^{i(k_y y+k_z z)}).
\end{align}
In this case, in order for $\Box_{4}\phi_{0}=0$ to be satisfied, we must have:
\begin{align}
\Box_2\varphi_{00}&=0
\nonumber \\
(\Box_2-m^2)\varphi_{01}&=0,
\end{align}
where $m^2=k_y^2+k_z^2$. Let us expand $\varphi_{01}$ in orders of $m^2a^2$ so that:
\begin {align}
&\varphi_{01}=\varphi_{01}^{(0)}+\varphi_{01}^{(1)}+...\,\, .
\end{align}
Matching powers of $m^2a^2$ we see that if $\varphi_{01}^{(0)}$ satisfies $\Box_2\varphi_{01}^{(0)}=0$, then the equation that is to be satisfied by $\varphi_{01}^{(1)}$ is:
\begin{align}
a^2\Box_2\varphi_{01}^{(1)}&=m^2a^2\varphi_{01}^{(0)}.
\label{cond1}
\end{align}
Anticipating that the leading modification, as in section
\ref{sec:non-spherical}, due to the deformation will be a shift in the
peak position of the wavepacket, we will make the ansatz up to terms of
order $r/a$: 
\begin{align}
&\varphi_{00}=Af(\tau_+)+Af(\tau_-)
\nonumber \\
&\varphi_{01}^{(0)}=-\frac{Ar_1}{a}[f'(\tau_+)+f'(\tau_-)],
\end{align}
where $\tau_\pm\equiv\frac{t_\pm}{a}\equiv\frac{t\pm x}{a}$. 

\begin{figure}[htb]
\begin{center}
\includegraphics[width=8.0cm]{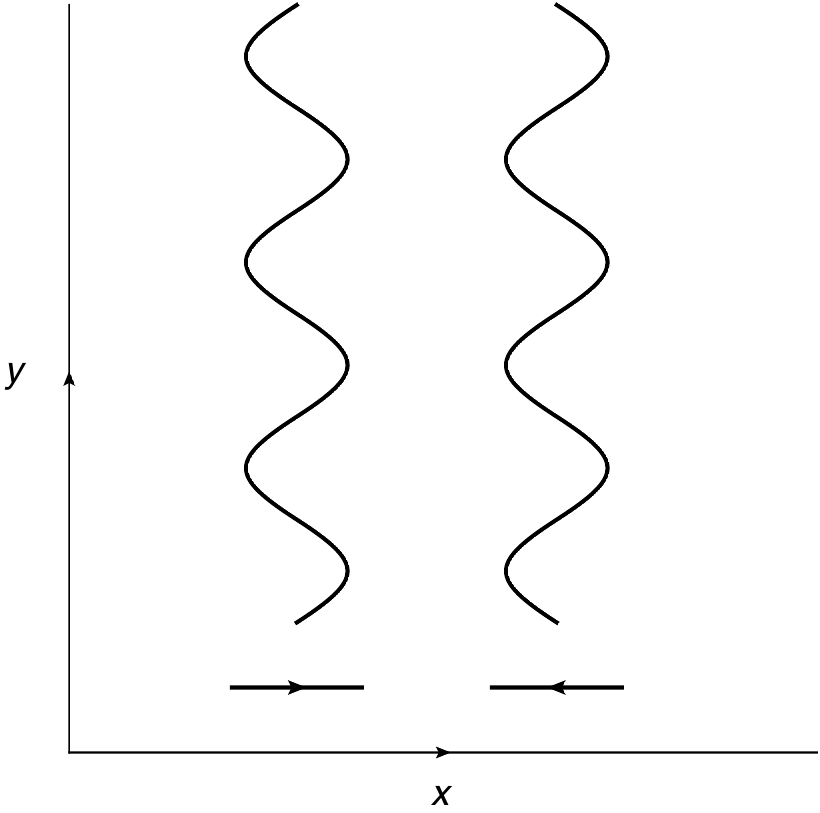}
\caption{A sketch showing the peak of $\phi_0$ in the $x$-$y$ plane for some time $t<0$.} 
\label{fig:p0}
\end{center}
\end{figure}

%Note that we can satisfy \eqref{cond1} by choosing:
%\begin{align}
%\varphi_{01}^{(1)}=-\frac{Am^2ar_1}{4}[&-\tau_+\int_{\tau0}^{\tau-}(f'(\tau'_-)-f(\tau'_-))d\tau'_-
%\nonumber \\
%&+\tau_-\int_{\tau0}^{\tau+}(f'(\tau'_+)+f(\tau'_+))d\tau'_+]
%\end{align}
%\begin{align}
%\varphi_{01}^{(2)}=-Am^2r_1^2[&-\tau_+\int_{\tau0}^{\tau-}f'(\tau'_-)d\tau'_-
%\nonumber \\
%&+\tau_-\int_{\tau0}^{\tau+}f'(\tau'_+)d\tau'_+]
%\end{align}
%where $\tau_0$ is an IR cutoff such to ensure that the massive wave modes do not spread out too far before coming together. Note that the above terms are of order $m^2a^2(r_1/a)$ and $m^2a^2(r_1^2/a^2)$. As in section 3, such terms represent the tail of the wavepacket, i.e., a small change in the overall behaviour. We will approximate them as negligible in the following analysis, leaving us with just massless modes, so we will not need an IR cutoff from here on out. 
We begin our analysis to determine if there is a change in the classicalization radius, $r_*$ by calculating the deformed $\phi_1$ to order $r_1/a$. Then we have for $\phi_0$ in this approximation:
\begin{align}
\phi_{0}(\tau_+,\tau_-,y,z)
&\approx\frac{A}{\sqrt{\mathcal{A}_2}}[f(\bar{\tau}_+)+f(\bar{\tau}_-)]
\nonumber \\
&\equiv\frac{A}{\sqrt{\mathcal{A}_2}}f(\bar{\tau}_+,\bar{\tau}_-),
\end{align}
where 
$\bar{\tau}_\pm=\frac{t_\pm -r_1\cos{(k_\perp\cdot x_\perp)}}{a}$.
Note that $f(\bar{\tau}_+)$ is a wave peaked at
$x=-t+r_1\cos{(k_\perp\cdot x_\perp)}$ and $f(\bar{\tau}_-)$ is a wave
peaked at $x=t-r_1\cos{(k_\perp\cdot x_\perp)}$. A cartoon of $\phi_0$
at a time slice $t<0$ is shown in figure \ref{fig:p0}.
We are interested in the region where the two plane waves are bent away
from each other, which corresponds to a deformation that brings us from
the perfect planar case closer to that of the spherical case. For our
ansatz, this corresponds to the region where 
$r_1 \cos{(k_\perp\cdot x_\perp)}$ is positive. We expect that $r_*$
will tend to increase in this region as we increase $r_1$.

We now need to find $\phi_1$ using the equation of motion:
\begin{align}
\Box_4\phi=L_*^4\partial^{\mu}(\partial_{\mu}\phi(\partial_{\nu}\phi)^2),
\end{align}
for which the first order correction is:
\begin{align}
\Box_4\phi_1=L_*^4\partial^{\mu}(\partial_{\mu}\phi_0(\partial_{\nu}\phi_0)^2).
\label{EOM}
\end{align}
One such contribution to this order, $\phi_{10}$, is obtained by substituting $\frac{1}{\sqrt{\mathcal{A}_2}}\varphi_{00}$ for $\phi_0$ above.:
\begin{align}
\Box_4\phi_{10}=\frac{8L_*^4A^3}{a^4\mathcal{A}_2^{3/2}}b(\tau_+,\tau_-),
\end{align}
where we have defined
$b(\tau_+,\tau_-)=f''(\tau_+)f'(\tau_-)^2+f''(\tau_-)f'(\tau_+)^2$ for
later convenience. It is not hard to show that the solution to the above
differential equations in the limit $m^2a^2\to 0$ is: 
\begin{align}
\phi_{10}=-\frac{2L_*^4A^3}{a^2\mathcal{A}_2^{3/2}}[&g(\tau_+,\tau_-)+\mathcal{O}(m^2a^2)],
\end{align}
where the function $g(\tau_+,\tau_-)$ is given by:
\begin{align}
&g(\tau_+,\tau_-)=\int^{\tau_+}_{-\infty}d\tau'_+\int^{\tau_-}_{-\infty}d\tau'_-b(\tau'_+,\tau'_-).
\end{align}
This solution is just the one presented in the previous section. In the following,  just as in the case for $\phi_0$, we will find the contribution to $\phi_1$ up to order $\frac{r_1}{a}$. We will denote this contribution
 $\phi_{11}^{(0)}$. 
 
$\phi_{11}^{(0)}$ can be found by considering the RHS of \eqref{EOM} with linear contributions from $\frac{1}{\sqrt{\mathcal{A}_2}}\varphi_{01}^{(0)}\cos{(k_\perp\cdot x_\perp)}$ and quadratic contributions from $\frac{1}{\sqrt{\mathcal{A}_2}}\varphi_{00}$. In this case:
\begin{align}
\Box_4\phi_{11}^{(0)}=-\frac{8L_*^4A^3r_1}{a^5\mathcal{A}_2^{3/2}}&\cos{(k_\perp\cdot x_\perp)}[\partial_{\tau+}b(\tau_+,\tau_-)
\nonumber \\
&+\partial_{\tau-}b(\tau_+,\tau_-)].
\label{eqbox11}
\end{align}
The solution to the above differential equation in the limit 
$m^2a^2\to 0$ is easily obtained: 
\begin{align}
\phi_{11}^{(0)}&=\frac{2L_*^4A^3r_1}{a^3\mathcal{A}_2^{3/2}}\cos{(k_\perp\cdot x_\perp)}[\partial_{\tau+}g(\tau_+,\tau_-)
\nonumber \\
&+\partial_{\tau-}g(\tau_+,\tau_-)+\mathcal{O}(m^2a^2)].
\label{eqphi11}
\end{align}
Adding up both contributions, we get that:
\begin{align}
\phi_1=-\frac{2L_*^4A^3}{a^2\mathcal{A}_2^{3/2}}g(\bar{\tau}_+,\bar{\tau}_-).
\end{align}
From this we get:
\begin{align}
\frac{\phi_1}{\phi_0}=-\frac{2L_*^4A^2}{a^2\mathcal{A}_2}[\frac{g(\bar{\tau}_+,\bar{\tau}_-)}{f(\bar{\tau}_+,\bar{\tau}_-)}+\mathcal{O}(\frac{r_1^2}{a^2})+\mathcal{O}(m^2a^2)].
\end{align}
Thus, we conclude that we need to consider the $\mathcal{O}(m^2a^2\frac{r_1}{a})$ corrections in order to see effects on $r_*$.

The only order $(m^2a^2\frac{r_1}{a})$ correction to $\phi_0$ will be to $\varphi_{01}$ since our expression for $\varphi_{00}$ satisfies its free field equation of motion exactly. In order for \eqref{cond1} to be satisfied, we must have:
\begin{align}
\varphi_{01}^{(1)}&=\frac{Ar_1m^2a}{4}[(\tau_--\tau_0)\int^{\tau_+}_{\tau_0}d\tau'_+f'(\tau'_+)
\nonumber \\
&+(\tau_+-\tau_0)\int^{\tau_-}_{\tau_0}d\tau'_-f'(\tau'_-)]
\nonumber \\
&=\frac{Ar_1m^2a}{4}[(\tau_--\tau_0)(f(\tau_+)-f(\tau_0))
\nonumber \\
&+(\tau_+-\tau_0)(f(\tau_-)-f(\tau_0))]
\nonumber \\
&\equiv \frac{Ar_1m^2a}{4}w(\tau_+,\tau_-),
\end{align}
where $\tau_0$ is an IR cutoff introduced to ensure that the massive wave modes do not spread out too far before coming together. We have also defined the function $w$ for later convenience. Our expression for $\phi_0$ is now:
\begin{align}
\phi_0=\frac{A}{\sqrt{\mathcal{A}_2}}[&f(\bar{\tau}_+,\bar{\tau}_-)-\frac{r_1m^2a}{4}\cos{(k_\perp\cdot x_\perp)}w(\tau_+,\tau_-)
\nonumber \\
&+\mathcal{O}(\frac{r_1^2}{a^2})+\mathcal{O}(m^4a^4)].
\end{align}
We now need to consider the order $m^2a^2(\frac{r_1}{a})$ corrections to $\phi_1$. There are actually only three such contributions. One will come from the fact that our solution for $\phi_{11}$ \eqref{eqphi11} to the differential equation \eqref{eqbox11} neglected terms of order $m^2a^2(\frac{r_1}{a})$ (we will denote this contribution $\phi_{11}^{(1b)}$). Another contribution will be found by considering the RHS of \eqref{EOM} to linear order in $\varphi_{01}^{(1)}$ (we will denote this contribution $\phi_{11}^{(1a)}$). The last contribution will come from that fact we now need to consider a term of order $m^2a^2(\frac{r_1}{a})$ that comes from the linear contribution of $\varphi_{01}^{(0)}$ in the RHS of \eqref{EOM} (we will denote this contribution $\phi_{11}^{(1c)}$). All other corrections are of order $\frac{r_1^2}{a^2}$ or $m^4a^4$. In order to get $\phi_{11}^{(1b)}$ all we have to do is note that if we make the ansatz:
\begin{align}
\phi_{11}^{(0)}&=\frac{2L_*^4A^3r_1}{a^3\mathcal{A}_2^{3/2}}\cos{(k_\perp\cdot x_\perp)}
\nonumber \\
&\int^{\tau_+}_{-\infty}d\tau'_+\int^{\tau_-}_{-\infty}d\tau'_-h_{b}(\tau'_+,\tau'_-)
\nonumber \\
&-\frac{L_*^4A^3r_1m^2}{2a\mathcal{A}_2^{3/2}}\cos{(k_\perp\cdot x_\perp)}
\nonumber \\
&\int^{\tau_+}_{\tau_0}d\tau'_+\int^{\tau'_+}_{-\infty}d\tau''_+\int^{\tau_-}_{\tau_0}d\tau'_-\int^{\tau'_-}_{-\infty}d\tau''_-h_{b}(\tau''_+,\tau''_-)
\nonumber \\
&+\mathcal{O}(\frac{r_1^2}{a^2})+\mathcal{O}(m^4a^4),
\end{align}
where
\begin{align}
h_{b}(\tau_+,\tau_-)=\partial_{\tau+}b(\tau_+,\tau_-)+\partial_{\tau-}b(\tau_+,\tau_-),
\end{align}
then \eqref{eqbox11} is satisfied to order $m^2a^2(\frac{r_1}{a})$.
Thus:
\begin{align}
\phi_{11}^{(1b)}&=-\frac{L_*^4A^3r_1m^2}{2a\mathcal{A}_2^{3/2}}\cos{(k_\perp\cdot x_\perp)}
\nonumber \\
&\int^{\tau_+}_{\tau_0}d\tau'_+\int^{\tau'_+}_{-\infty}d\tau''_+\int^{\tau_-}_{\tau_0}d\tau'_-\int^{\tau'_-}_{-\infty}d\tau''_-h_{b}(\tau''_+,\tau''_-).
\end{align}
In order to find $\phi_{11}^{(1a)}$ we need to solve \eqref{EOM} with linear order contributions from  $\varphi_{01}^{(1)}$ on the RHS. In this case, the equation becomes:
\begin{align}
\Box_4\phi_{11}^{(1a)}&=\frac{2L_*^4A^3r_1m^2}{a^3\mathcal{A}_2^{3/2}}\cos{(k_\perp\cdot x_\perp)}[(\tau_+-\tau_0)f'(\tau_+)^2f''(\tau_-)
\nonumber \\
&+(\tau_--\tau_0)f'(\tau_-)^2f''(\tau_+)
\nonumber \\
&+2f'(\tau_-)f''(\tau_+)((\tau_+-\tau_0)f'(\tau_-)+f(\tau_+)-f(\tau_0))
\nonumber \\
&+2f'(\tau_+)f''(\tau_-)((\tau_--\tau_0)f'(\tau_+)+f(\tau_-)-f(\tau_0))]
\nonumber \\
&\equiv \frac{2L_*^4A^3r_1m^2}{a^3\mathcal{A}_2^{3/2}}\cos{(k_\perp\cdot x_\perp)}h_{a}(\tau_+,\tau_-).
\end{align}
The solution up to the order we seek is:
\begin{align}
\phi_{11}^{(1a)}&=-\frac{L_*^4A^3r_1m^2}{2a\mathcal{A}_2^{3/2}}\cos{(k_\perp\cdot x_\perp)}
\nonumber \\
&\int^{\tau_+}_{\tau_0}d\tau'_+\int^{\tau_-}_{\tau_0}d\tau'_-h_{a}(\tau'_+,\tau'_-).
\end{align}
The final correction of this order comes from the order $m^2a^2(\frac{r_1}{a})$ term that arises from linear contributions from $\varphi_{01}^{(0)}$ on the RHS of \eqref{EOM}:
\begin{align}
\Box_4\phi_{11}^{(1c)}&=\frac{2L_*^4A^3r_1m^2}{a^3\mathcal{A}_2^{3/2}}\cos{(k_\perp\cdot x_\perp)}
\nonumber \\
&[2(f'(\tau_+)+f'(\tau_-))]f'(\tau_+)f'(\tau_-)
\nonumber \\
&\equiv \frac{2L_*^4A^3r_1m^2}{a^3\mathcal{A}_2^{3/2}}\cos{(k_\perp\cdot x_\perp)}h_{c}(\tau_+,\tau_-).
\end{align}
So we have that the order $m^2a^2(\frac{r_1}{a})$ correction to $\phi_1$ is:
\begin{align}
\phi_{11}^{(1)}&=\phi_{11}^{(1a)}+\phi_{11}^{(1b)}+\phi_{11}^{(1c)}
\nonumber \\
&=-\frac{L_*^4A^3r_1m^2}{2a\mathcal{A}_2^{3/2}}\cos{(k_\perp\cdot x_\perp)}
\nonumber \\
&[\int^{\tau_+}_{\tau_0}d\tau'_+\int^{\tau'_+}_{-\infty}d\tau''_+\int^{\tau_-}_{\tau_0}d\tau'_-\int^{\tau'_-}_{-\infty}d\tau''_-h_{b}(\tau''_+,\tau''_-)
\nonumber \\
&+\int^{\tau_+}_{\tau_0}d\tau'_+\int^{\tau_-}_{\tau_0}d\tau'_-(h_{a}(\tau'_+,\tau'_-)+h_{c}(\tau'_+,\tau'_-))]
\nonumber \\
&\equiv -\frac{L_*^4A^3r_1m^2}{2a\mathcal{A}_2^{3/2}}\cos{(k_\perp\cdot x_\perp)}q(\tau_+,\tau_-).
\end{align}
So, after adding in this contribution, we have:
\begin{align}
\phi_1=-\frac{2L_*^4A^3}{a^2\mathcal{A}_2^{3/2}}g(\bar{\tau}_+,\bar{\tau}_-)[&1+\frac{m^2r_1a}{4}\cos{(k_\perp\cdot x_\perp)}\frac{q(\tau_+,\tau_-)}{g(\bar{\tau}_+,\bar{\tau}_-)}
\nonumber \\
&+\mathcal{O}(\frac{r_1^2}{a^2})+\mathcal{O}(m^4a^4)].
\end{align}
We thus get that:
\begin{align}
\frac{\phi_1}{\phi_0}&=-\frac{2L_*^4A^2}{a^2\mathcal{A}_2}\frac{g(\bar{\tau}_+,\bar{\tau}_-)}{f(\bar{\tau}_+,\bar{\tau}_-)}
\nonumber \\
&[1+\frac{m^2r_1a}{4}\cos{(k_\perp\cdot x_\perp)}(-\frac{w(\tau_+,\tau_-)}{f(\bar{\tau}_+,\bar{\tau}_-)}+\frac{q(\tau_+,\tau_-)}{g(\bar{\tau}_+,\bar{\tau}_-)})
\nonumber \\
&+\mathcal{O}(\frac{r_1^2}{a^2})+\mathcal{O}(m^4a^4)].
\end{align}

What we are interested in is the sign of the quantity: $(-\frac{w(\tau_+,\tau_-)}{f(\bar{\tau}_+,\bar{\tau}_-)}+\frac{q(\tau_+,\tau_-)}{g(\bar{\tau}_+,\bar{\tau}_-)})$. It is clear that $f(\bar{\tau}_+,\bar{\tau}_-)$ is always positive for $f(t)=e^{-t^2}$. Writing $g(\bar{\tau}_+,\bar{\tau}_-)$ explicitly for this ansatz for $f(t)$, we have:
\begin{align}
g(\bar{\tau}_+,\bar{\tau}_-)&=\frac{1}{2}e^{2(\bar{\tau}_+^2+\bar{\tau}_-)^2}[4e^{\bar{\tau}_-^2}\bar{\tau}_-\bar{\tau}_++4e^{\bar{\tau}_+^2}\bar{\tau}_-\bar{\tau}_+
\nonumber \\
&+\sqrt{2\pi}e^{2\bar{\tau}_-^2+\bar{\tau}_+^2}\bar{\tau}_+(-2+\text{erfc}(\sqrt{2}\bar{\tau}_-)
\nonumber \\
&+\sqrt{2\pi}e^{2\bar{\tau}_+^2+\bar{\tau}_-^2}\bar{\tau}_-(-2+\text{erfc}(\sqrt{2}\bar{\tau}_+)].
\end{align}
It can be seen that $g(\bar{\tau}_+,\bar{\tau}_-)>0$ for $\bar{\tau}_+,\bar{\tau}_-<0$, which is the region we are interested in. Since:
\begin{align}
&-\frac{w(\tau_+,\tau_-)}{f(\bar{\tau}_+,\bar{\tau}_-)}+\frac{q(\tau_+,\tau_-)}{g(\bar{\tau}_+,\bar{\tau}_-)}
\nonumber \\
&=\frac{1}{f(\bar{\tau}_+,\bar{\tau}_-)g(\bar{\tau}_+,\bar{\tau}_-)}[-w(\tau_+,\tau_-)g(\tau_+,\tau_-)
\nonumber \\
&+q(\tau_+,\tau_-)f(\tau_+,\tau_-)+\mathcal{O}(\frac{r_1}{a})].
\end{align}
We are interested in the sign of the quantity:
\begin{align}
\sigma(\tau_+,\tau_-)=-w(\tau_+,\tau_-)g(\tau_+,\tau_-)+q(\tau_+,\tau_-)f(\tau_+,\tau_-).
\end{align}

\begin{figure}[htb]
\begin{center}
\includegraphics[width=8.0cm]{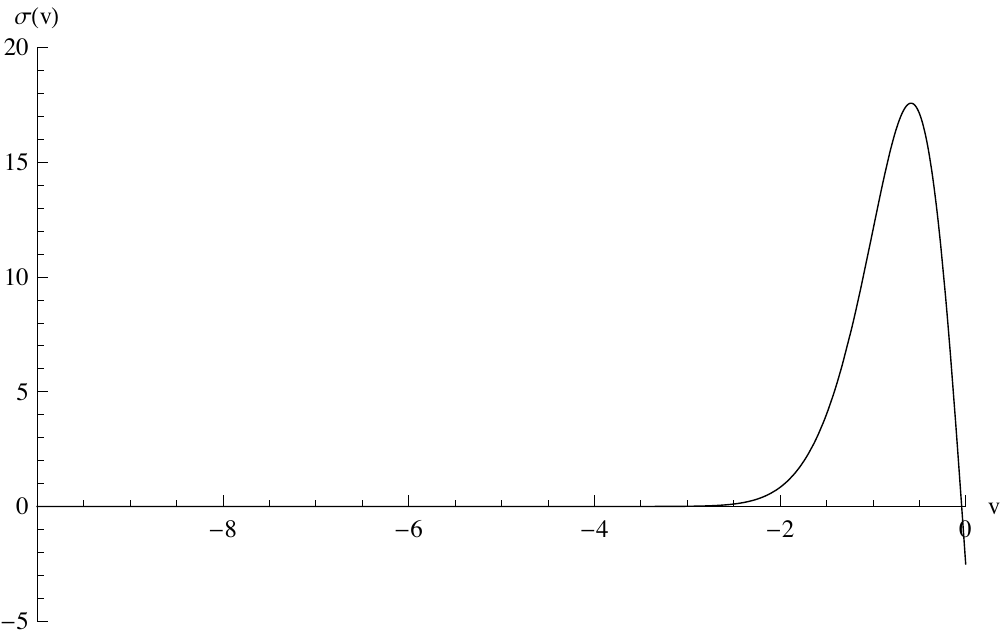}
\caption{$\sigma(\tau_+=v,0)$ as a function of $v$. $\tau_0=-10$ was chosen, but any $\tau_0$ much smaller than the $\tau_-$, $\tau_+$ where the interaction takes place defines an appropriate boundary condition.} 
\label{fig:sigma}
\end{center}
\end{figure}

Let us plot $\sigma(\tau_+=v,0)$ in order to find the sign. We see in figure \ref{fig:sigma} that $\sigma(\tau_+=v,0)$ remains positive until $v>>-1$. As we saw previously, the interaction length for the perfect planar case is of order $a$ (which corresponds to the region where $v\sim -1$), so we see that increasing $r_1$ tends to increase $r_*$ in the region where $r_1 \cos{(k_\perp\cdot x_\perp)}$ is positive. This is what we expected.

\section{UV sensitivity}
\label{sec:UVsensitivity}

We have so far explored a model with the Lagrangian:

\begin{align}
\mathcal{L}=X+L_*^4X^2,
\end{align}
where $X=-\frac{1}{2}\partial^\mu\phi\partial_\mu\phi$ is the standard kinetic term. Let us generalize this so as to include the possibility of a higher order non-linear term:
\begin{align}
\mathcal{L}=X+L_*^{4n-4}X^n.
\end{align}
For the planarly symmetric case, we have for the first order EOM:
\begin{align}
\Box_2\varphi_1=-nL_*^{4n-4}
 \partial^\mu \left[\varphi_0 \partial_\mu(X_0^{n-1})\right],
\label{Order1EOM}
\end{align}
where
$X_0=-\frac{1}{2{\cal A}_2}\partial^{\mu}\varphi_0\partial_{\mu}\varphi_0$. If we choose:
\begin{align}
\varphi_0=Af(\tau_+)+Af(\tau_-).
\end{align}
Then:
\begin{align}
X_0=\frac{2A^2}{{\cal A}_2l^2}f'(\tau_+)f'(\tau_-).
\end{align}
In this case, \eqref{Order1EOM} simplifies to:
\begin{align}
\Box_2\varphi_1=&n(n-1)2^nA^{2n-1}{\cal A}_2^{-n+1}L_*^{4n-4}l^{-2n}
\nonumber \\
&[f'(\tau_+)^nf'(\tau_-)^{n-2}f''(\tau_-)
\nonumber \\
+&f'(\tau_-)^nf'(\tau_+)^{n-2}f''(\tau_+)].
\end{align}
The retarded solution to the above equation is:
\begin{align}
\varphi_1=-A\epsilon
&\int_{-\infty}^{\tau+}d\tau'_+\int_{-\infty}^{\tau-}d\tau'_-[f'(\tau'_+)^nf'(\tau'_-)^{n-2}f''(\tau'_-)
\nonumber \\
&+f'(\tau'_-)^nf'(\tau'_+)^{n-2}f''(\tau'_+)].
\end{align}
where 
\begin{align}
\epsilon=n(n-1)2^{n-2}A^{2n-2}{\cal A}_2^{-n+1}L_*^{4n-4}l^{2-2n}.
\end{align}

\begin{figure}[htb]
\begin{center}
\includegraphics[width=8.0cm]{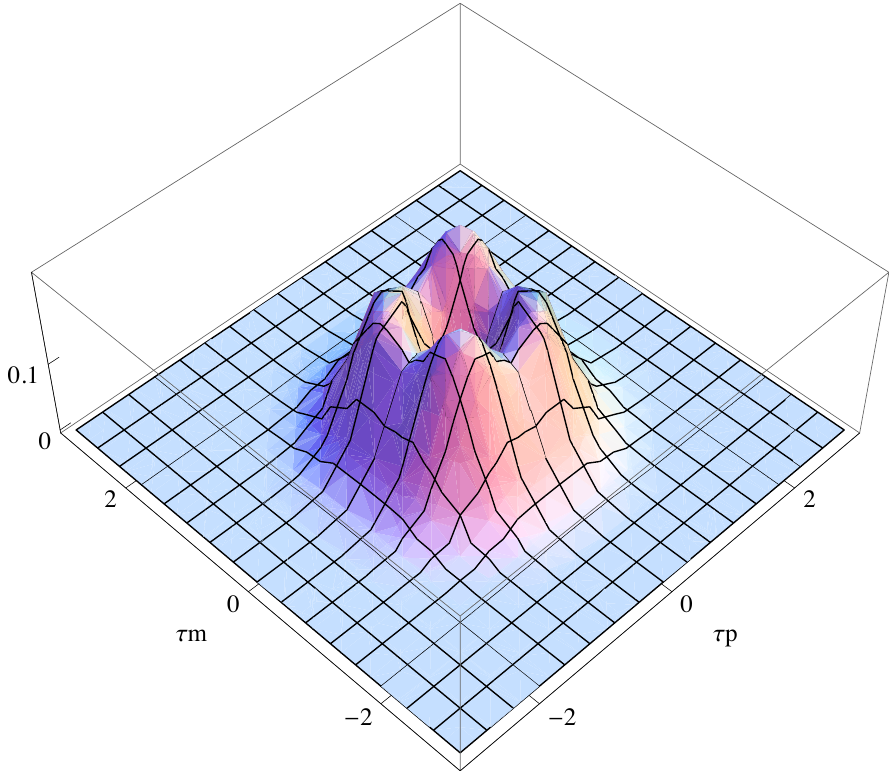}
\caption{The plot of $\varphi_1/(-A\epsilon)$ as a function of 
$\tau_{\pm}\equiv t_{\pm}/a$ for $f(\tau)=e^{-\tau^2}$, $n=3$.
} 
\label{fig:phi13}
\end{center}
\end{figure}

It can be seen by comparing figure \ref{fig:phi1}
($\varphi_1/(-A\epsilon)$ for the $n=2$ case) to figure \ref{fig:phi13}
($\varphi_1/(-A\epsilon)$ for the $n=3$ case) that the overall profile
of $\phi_1$ for odd and even values of $n$ can be quite different. It
should also be noted that we find relatively modest dependence of
the profiles of two even (or two odd) $\phi_1$ on $n$.

In order to compare the behavior of $\varphi_{1}$ for various values of
$n$, let us plot the quantity:  
\begin{align}
\xi(v)=\frac{\varphi_1(\tau_+=v,\tau_-=0)}{\text{Max}_v[\varphi_1(\tau_+=v,\tau_-=0)]}.
\end{align}
The function $\xi(v)$ is the normalized peak amplitude of the right
moving plane wave $f(\tau_-)$. We normalize by
$\text{Max}_v[\varphi_1(\tau_+=v,\tau_-=0)]$, the maximum value of
$\varphi_1(\tau_+=v,\tau_-=0)$ in the region $-\infty<v\le0$, so that
the peak value of $\xi(v)$ is 1 regardless of what $n$ we choose. Note
that since we set $\tau_-=0$, $v=\frac{2t}{a}=\frac{2x}{a}$. Thus, the
right moving plane wave evolves from $v=-\infty$ to $v=0$, where it will
meet the left moving plane wave. 

\begin{figure}[htb]
\begin{center}
\includegraphics[width=8.0cm]{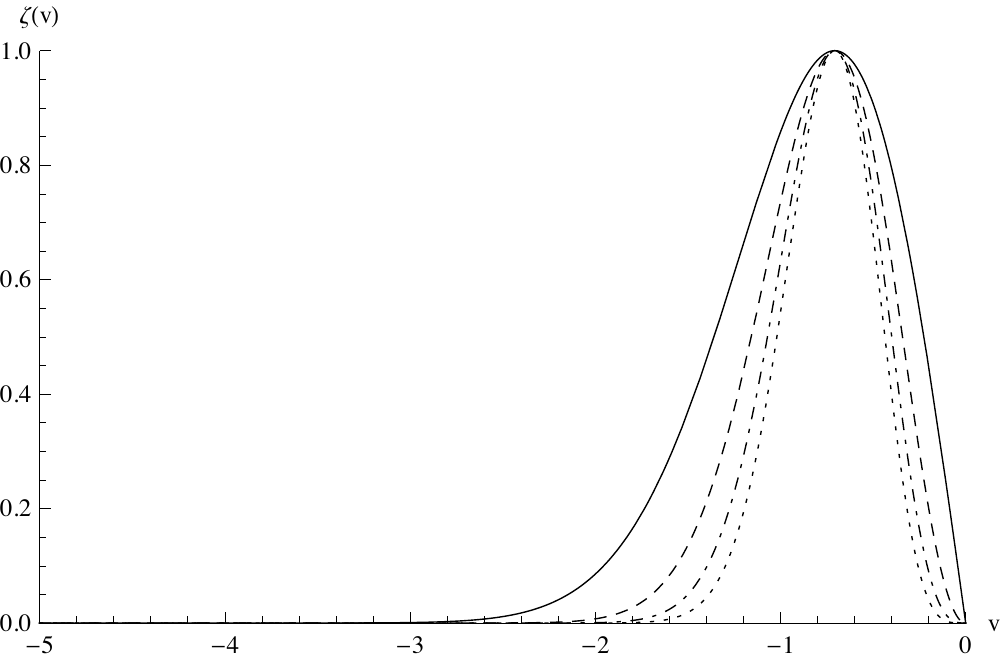}
\caption{The plot of $\xi(v)$ for $n=2$ (solid line), $n=3$ (dashed line), $n=4$ (dot-dashed line), $n=5$ (dotted line).} 
\label{fig:allp}
\end{center}
\end{figure}

We see in figure \ref{fig:allp} that $\xi(v)$ reaches its peak value at
the value of $v=\frac{\sqrt{2}}{2}$ for all plots shown. Thus, the
interaction length is $\sim a$ regardless of what value we choose for
$n$. It is also apparent that as we increase $n$, there is a tendency
for $\xi(v)$ to become more sharply peaked, so $\phi_1$ will become
significant at somewhat smaller values of $x$ as we increase $n$. Hence
we can conclude that increasing $n$ will not cause classicalization to
occur for planarly symmetric waves. Moreover, the shape of the scattered
wave is UV sensitive. 

\section{Conclusion}
\label{sec:conclusion}

We have pointed out that a field theory that exhibits the classicalization
phenomenon for perfect spherical symmetry ceases to do so when the
spherical symmetry is significantly relaxed. We have shown that the
classicalization radius tends to decrease in a region where a shell made
of the field is flattened and that in the planar limit, the system never
classicalizes before reaching sub-cutoff lengths. We have also seen that
the shape of the scattered wave is UV sensitive. These considerations
point towards the conclusion that classicalization does not serve as
UV-completion for the class of non-renormalizable theories.

\begin{acknowledgments} 
The work of RA and RS was supported in part by the US Department of Energy.
 Part of this work was done during S.M.'s visit at MCTP, University of 
 Michigan, Ann Arbor. He would like to express his gratitude to the
 center for warm hospitality. S.M. acknowledges the support by
 Grant-in-Aid for Scientific Research 17740134, 19GS0219, 21111006,
 21540278, by Japan-Russia Research Cooperative Program, and by WPI
 Initiative, MEXT, Japan. 
\end{acknowledgments}

\end{document}